\documentclass[a4paper]{article}
\usepackage{latexcad}
\usepackage{amsmath,amssymb}
\numberwithin{equation}{section}
\def\be#1\ee{\begin{equation}#1\end{equation}}

\newcommand{\al}{\alpha}

\newcommand{\bet}{\beta}

\newcommand{\om}{\omega}

\newcommand{\G}{\Gamma}

\newcommand{\si}{\sigma}

\newcommand{\la}{\lambda}

\newcommand{\eps}{\varepsilon}
\newcommand{\dz}{\wedge}

\newcommand{\ba}{\begin{array}}
\newcommand{\ea}{\end{array}}
\newcommand{\beq}{\begin{eqnarray}}
\newcommand{\eeq}{\end{eqnarray}}
\newtheorem{lm}{Lemma}
\newtheorem{thee}{Theorem}
\newtheorem{proo}{Proposition}
\newtheorem{co}{Corollary}
\newtheorem{rem}{Remark}
\newtheorem{deff}{Definition}
\newcommand{\bd}{\begin{deff}}
\newcommand{\ed}{\end{deff}}

\newcommand{\bl}{\begin{lm}}
\newcommand{\el}{\end{lm}}
\newcommand{\bp}{\begin{proo}}
\newcommand{\ep}{\end{proo}}
\newcommand{\bt}{\begin{thee}}
\newcommand{\et}{\end{thee}}
\newcommand{\bc}{\begin{co}}
\newcommand{\ec}{\end{co}}
\newcommand{\brm}{\begin{rem}}
\newcommand{\erm}{\end{rem}}

\newcommand{\der}{{\rm d}}
\begin{document}

\thispagestyle{empty}

\title {INTRINSIC GEOMETRY OF A NULL HYPERSURFACE 
\footnote{Research supported 
by 
Komitet Bada\'{n} Naukowych (Grant nr 2 P03B 060 17), the 
Erwin Schr\"{o}dinger International Institute for Mathematical Physics and the London Mathematical Society.}\\ 
\vskip 1.truecm
{\small {\sc Pawe\l  ~Nurowski}} 
\footnote{{\small{\small {\tt e-mail: nurowski@fuw.edu.pl
}}}}\\
\vskip -0.3truecm
{\small {\it Instytut Fizyki Teoretycznej}}\\
\vskip -0.3truecm
{\small {\it Uniwersytet Warszawski}}\\
\vskip -0.3truecm
{\small {\it ul. Ho\.za 69, Warszawa, Poland }}\\
{\small { and}}\\
{\small {\sc David C. Robinson}}
\footnote{{\small{\small {\tt e-mail: david.c.robinson@kcl.ac.uk
}}}}\\
\vskip -0.3truecm
{\small {\it Department of Mathematics}}\\
\vskip -0.3truecm
{\small {\it King's College London}}\\
\vskip -0.3truecm
{\small {\it Strand, London WC2R 2LS, U.K.}}
}
\author{\mbox{}}
\maketitle
\begin{abstract}
We apply Cartan's method of equivalence 
to construct invariants of a given 
null hypersurface in a Lorentzian space-time. This enables us to fully 
classify the internal geometry of such surfaces and hence 
solve the local equivalence problem for null hypersurface structures in 
4-dimensional Lorentzian space-times. 
\end{abstract}

\newpage
\noindent
\section{Introduction}
The study of the structure of null hypersurfaces in four-dimensional 
space-time has played a key role in the development of general relativity 
and the mathematics and physics of gravitation. For example detailed 
investigations of null hypersurfaces have been neccessary in order to 
understand the causal structure of space-times, black holes,
asymptotically flat systems and gravitational waves. In this paper we
shall study and classify the internal geometries of null
hypersurfaces.  \\

\noindent
In J. L. Synge's festshrift volume \cite{bi:Synge} Roger Penrose
distinguished three types of geometries which a null hypersurface 
$\cal N$ in 4-dimensional space-time $({\cal M}, g)$, acquires from the 
ambient Lorentzian geometry. 
These geometries are associated with the following geometrical structures
that are defined on $\cal N$:
\begin{itemize}
\item[i)] the degenerate metric $g_{|\cal N}$
\item[ii)] the concept of an affine parameter along each of the null geodesics
from the 2-parameter family ruling $\cal N$
\item[iii)] the concept of parallel transport for tangent vectors to 
$\cal N$ along each of the null geodesics.
\end{itemize}
Penrose's three geometries are then:
\begin{itemize}
\item[I] - the geometry of structure i),
\item[II] - the geometry of structures i) and ii),
\item[III] - the geometry of structures i) and iii).
\end{itemize}
In this paper we study the weakest of the Penrose
geometries, I - geometry. In this context it is convenient to use
the following definition of a null hypersurface. 
\bd
A null hypersurface in an oriented and time oriented space-time
$({\cal M},g)$ is a 3-dimensional submanifold $\cal N$ in $\cal M$ which is 
such that the restricted metric $g_{|\cal N}$ is degenerate.
\ed
It follows that if the metric $g$ has signature $(+ + + -)$ then the signature
of $g_{|\cal N}$ is $(+ + 0)$.\\

\noindent
Since in this paper we are investigating only 
Penrose's I - geometry it is convenient to introduce the concept of a
null hypersurface structure (NHS).
\bd
A null hypersurface structure (NHS) is a 3-dimensional oriented
manifold $\cal N$ equipped with a degenerate metric $h$ of signature
$(+ + 0)$.
\ed
Note that any null hypersurface in space-time defines an NHS.\\

\noindent
Our main aim in this paper is to solve the local equivalence problem for NHS's by using the methods developed by E. Cartan (a helpful modern exposition has been presented by Olver \cite{bi:Olver}). Consequently, we 
introduce the following definitions. 
\bd
Two NHSs $({\cal N}_1,h_1)$ and $({\cal N}_2,h_2)$ are (locally)
equivalent if and only if there exists a (local) diffeomorphism 
$$\psi:{\cal N}_1\to {\cal N}_2$$
such that 
$$\psi^* h_2=h_1.$$
\ed
\bd
A diffeomorphism $\psi$ is called a symmetry of a NHS $({\cal N},h)$ iff 
$$\psi^*h=h.$$ 
\ed
The local version of this definition is:
\bd
A vector field $X$ on $\cal N$ is an infinitesimal symmetry of a NHS
$({\cal N},h)$ iff 
$${\cal L}_X h=0.$$
\ed
It follows that the set of all infinitesimal symmetries naturally has the 
structure of a Lie algebra; the Lie algebra of symmetries of the NHS.  \\

\noindent
Cartan's method for dealing with local equivalence problems is
explained fully in Reference \cite{bi:Olver}. Here we outline, using
the notation of that book, only the basic ideas and the definitions
which are particularly important in the sequel. \\
Given a geometrical structure on a manifold, the first step is to redefine it in
terms of a coframe given up to certain transformations. 
More formally, with each geometrical structure one associates an
equivalence class of coframes 
$[(\om^i)]=[(\om^1, \om^2,...,\om^n)]$, with an equivalence relation
such that a coframe $(\om^i)$ is equivalent to a coframe
$(\tilde{\om}^i)$ if and only if there exists a function $(a^i_{~j})$ 
with values in a Lie subgroup $G$ of ${\bf GL}(n,{\bf R})$ such that 
$$\tilde{\om}^i=a^i_{~j}\om^j.$$ The group $G$ is totally
characterized by the geometrical structure considered. The structure is
then callled a $G$-structure.\\  
Now, the equivalence problem for two $G$-structures
translates into a problem of $G$-equivalence of the corresponding
coframes. Such a problem may have two, qualitatively different,
outcomes. Either a $G$-structure admits an infinite dimensional group of
local symmetries or it does not. 
In the first case the structure is said to be {\it
involutive} (or {\it in involution}).\\
Cartan's method provides an algorithm for determining whether a given
$G$-structure is in involution or not. The only operations one uses 
to determine this are differentiation and linear algebra. By
means of them, given a $G$-structure on an $n$-dimensional manifold,
one calculates 
the so called {\it degree of indeterminacy of the
$G$-structure}, $r^{(1)}$, and $n-1$ parameters
$s'_i$, called the {\it Cartan characters of the
$G$-structure}. The $n$th Cartan character $s'_n=r-s'_1-s'_2-...-s'_{n-1}$ 
is defined in terms of the
previous $n-1$ ones and the dimension $r$ of the group $G$. Cartan's result 
is that the
$G$-structure is in involution if the following equality 
$$1 s'_1+2 s'_2+3 s'_3+...+n s'_n=r^{(1)},$$
called the Cartan test, holds. Otherwise the $G$-structure is not
involutive.\\ 
In the involutive case Cartan's theorems give the solution to the
equivalence problem. In particular, they state that if $s'_k$
is the last nonvanishing Cartan character then the set of analytic
morphisms transforming a given $G$-structure to itself depends on
$s'_k$ analytic functions of $k$ variables.\\
If a $G$-structure is not involutive then, by means of
the techniques called {\it absorption}, {\it normalization} and {\it
prolongation}, one reduces the original $G$-equivalence problem for
the initial $G$-structure either
to an $\{e\}$-equivalence problem (possibly on a new manifold) or to a
new problem for a $G_1$-structure with a new group
$G_1$.\\ 
If it happens that the $G$-equivalence problem is reduced to
an $\{e\}$-problem then Cartan's theorems provide a simple 
method for the construction of all the invariants of the original
geometrical structure. If one ends with a new $G_1$-structure
then one asks whether it is involutive or whether it can be further
reduced to an $\{e\}$-structure. Cartan's theorems state that, under certain
regularity assumptions, each
$G$-equivalence problem, after a finite number $p$ of steps, will
finally reduce
either to an $\{e\}$-structure or to a $G_p$-structure, which is involutive. 
In the case of an $\{e\}$-structure Cartan gives, as before, a simple
method for constructing all the invariants of the original geometrical 
structure. In the involutive situation two cases may occur. In the first case, the
geometrical structure in question is unique modulo a local
$G$-equivalence. In the second case it is not, and Cartan's method produces
all the $G$-invariants. In both cases, if we denote the last nonvanishing 
Cartan character for the final involutive $G_p$ structure by $s'_k$, then 
the set of analytic
morphisms transforming a given $G$-structure to itself depends on
$s'_k$ analytic functions of $k$ variables.\\

\noindent 
It may be helpful at this point to note that a number of authors have
applied Cartan's method of equivalence to the Lorentzian 4-metric of
general relativity. Two examples of such work are given by references 
\cite{davcar1,davcar2}. We also note that geometrical structures, in
Cartan's sense, induced on null hypersurfaces in $n$-dimensional 
space-times have been investigated in reference \cite{bp1}. \\

\noindent
In the next section we outline the basic formalism which we use in this
paper. The following section contains a discussion of the G-structure
- a null hypersurface structure - to which we shall apply Cartan's
method. Section 4 contains the application of the equivalence
method. The main results of the paper are stated as a series of
propositions in that Section. The final Section contains some brief
comments on this work. In addition there are two appendices. Appendix A 
contains a description of null hypersurfaces structures with 
zero complex expansion, non-zero shear and three dimensional symmetry groups. 
Finally the results of
the classification computed in Section 4 are summarized
diagramatically in Appendix B.
\section{Space-time formalism} 
In this section we outline the space-time formalism which we shall use in 
this paper. In general we follow the conventions and notation of Reference 
\cite{bi:Kramer} and use the Newman-Penrose formalism as it is
presented in that reference.
Consider a 4-dimensional space-time ($\cal M$, $g$) with a metric $g$
of a signature $(+ + + -)$. Let 
$(e_1, e_2, e_3, e_4)=(m, \bar{m}, l, k)$  be a null tetrad on 
$\cal M$ with dual $(M, \bar{M}, L, K)$ so that $g=2(M\bar{M}-LK)$. For such 
a space-time the first Cartan structure equations can be written as follows:
\beq              
&\der M=-\G_{21}\dz M -\G_{23}\dz L-\G_{24}\dz K,\nonumber\\
&\der L=\G_{41}\dz M+\G_{42}\dz\bar{M}+\G_{43}\dz L,\label{car1}\\
&\der K=\G_{31}\dz M+\G_{32}\dz\bar{M}+\G_{34}\dz K,\nonumber
\eeq
where the connection 1-forms $\G_{ij}$, $(i,j=1,2,3,4)$ satisfy 
$\G_{ij}=-\G_{ji}$ and a change of indicies $1\leftrightarrow 2$ implies 
complex conjugation (e.g. $\Gamma_{13}$ is the complex conjugate of 
$\Gamma_{23}$). In terms of Newman-Penrose coefficients (see Ref. \cite{bi:Kramer} pp. 82-87) 
\beq
&\G_{41}=\sigma M+\rho\bar{M}+\tau L +\kappa K,\nonumber\\
&\G_{23}=\mu M + \lambda \bar{M}+\nu L +\pi K,\label{och}\\
&\frac{1}{2}(\G_{12}+\G_{34})=-\beta M-\alpha \bar{M}-\gamma L-\epsilon K.\nonumber
\eeq  
With this notation equations (\ref{car1}) read 
{\small 
\beq              
&\der M=(\al-\bar{\bet})M\dz \bar{M} +(\gamma-\bar{\gamma}-\mu)M\dz
L-\la\bar{M}\dz L+(\eps-\bar{\eps}+\bar{\rho})M\dz
K+\bar{\si}\bar{M}\dz K+(\bar{\tau}+\pi)L\dz K,\nonumber\\
&\der L=(\bar{\rho}-\rho)M\dz\bar{M}+(\bar{\al}+\bet-\tau)M\dz
L+(\al+\bar{\bet}-\bar{\tau})\bar{M}\dz L+\kappa K\dz
M+\bar{\kappa}K\dz\bar{M}+(\eps+\bar{\eps})K\dz L,\label{1car}\\
&\der K=(\bar{\mu}-\mu)M\dz\bar{M}+\bar{\nu}M\dz L+\nu\bar{M}\dz
L+(\bar{\pi}-\bet-\bar{\al})M\dz K+(\pi-\al-\bar{\bet})\bar{M}\dz
K+(\gamma+\bar{\gamma})K\dz L.\nonumber
\eeq}
The second Cartan structure equations are:
\beq
&\der\G_{23}=(\G_{12}+\G_{34})\dz \G_{23}+\Psi_4 \bar{M}\dz L+\Psi_3 (K\dz L-M\dz \bar{M})+
(\Psi_2 +\frac{1}{12}R)K\dz M\nonumber\\
&+\frac{1}{2}S_{33}M\dz L+\frac{1}{2}S_{32}(K\dz L+M\dz\bar{M})+
\frac{1}{2}S_{22}K\dz\bar{M},\nonumber\\
&\quad\nonumber\\
&\der\G_{14}=\G_{14}\dz(\G_{12}+\G_{34})+(-\Psi_2-\frac{1}{12}R)\bar{M}\dz L-
\Psi_1(K\dz L-M\dz \bar{M})-
\Psi_0 K\dz M\nonumber\\
&-\frac{1}{2}S_{11}M\dz L-\frac{1}{2}S_{41}(K\dz L+M\dz\bar{M})-
\frac{1}{2}S_{44}K\dz\bar{M},\label{car2}\\
&\quad\nonumber\\
&\frac{1}{2}(\der\G_{12}+\der\G_{34})=\G_{23}\dz\G_{14}-\Psi_3 \bar{M}\dz L-
(\Psi_2-\frac{1}{24}R) 
(K\dz L-M\dz \bar{M})-\Psi_1 K\dz M\nonumber\\
&-\frac{1}{2}S_{31}M\dz L-\frac{1}{4}(S_{12}+S_{34})(K\dz L+M\dz\bar{M})-
\frac{1}{2}S_{42}K\dz\bar{M},\nonumber
\eeq~\\
where we 
have introduced the (spinorial) Weyl tensor coefficients $\Psi_\mu$, 
$(\mu=0,1,2,3,4)$ and the traceless part of the Ricci tensor 
$S_{ij}=R_{ij}-g_{ij}R/4.$\\
It should also be noted here that the structure equations 
(\ref{car1}), (\ref{car2}) imply the Newman-Penrose equations (7.28)-(7.71) of Ref. \cite{bi:Kramer}.\\

\noindent
Since we shall be considering the geometry of null hypersurfaces ruled
by null geodesics with null tangent vector $k$ we shall also need to 
employ the null rotations which preserve the direction of $k$.\\
Let $A>0$, $\phi$ (real) and $z$ (complex) be functions on $\cal M$. Then 
the action of the Lorentz transformations on the null coframe, which preserve the $k$ direction is given by   
\beq
&M\to{\rm e}^{i\phi}~[~M + z L~],\nonumber\\
&L\to A^{-1}~L,\label{eq:Lork}\\
&K\to A ~[~K+z\bar{z}L+z\bar{M}+\bar{z} M~].\nonumber
\eeq
The corresponding transformations of the connection 
1-forms are given by:
\beq
&\G _{41}\to A^{-1}~{\rm e}^{-i\phi}~\G_{41}\nonumber\\
&\G _{23}\to A~{\rm e}^{i\phi}~[~\G_{23}+z(\G_{12}+\G_{34})-z^2\G_{14}-\der z~]
\label{eq:itr}\\
&\G _{12}+\G _{34}\to\G_{12}+\G_{34}+2z\G_{41}+\der\log A+i\der\phi.\nonumber
\eeq

\noindent
In subsequent sections we shall use these null frames and equations 
restricted to the null hypersurface $\cal N$. We shall not distinguish 
notationally 
between space-time and null hypersurface quantities but the distinction will 
be clear from the context.

\section{G-structure for null hypersurface structures (NHS's)}
To apply Cartan's method to a NHS $({\cal N},h)$ we first need to reformulate the definition of a NHS in terms of a co-frame defined, up to appropriate transformations, on $\cal N$. To achieve this, recall that each NHS
$({\cal N},h)$ has a  
metric tensor $h$ of  signature $(+ + 0)$. Thus, locally,  it can be 
written in the form $h=2M \bar{M}$, where $M$ is a  
complex-valued 1-form on $\cal N$, such that $M\dz\bar{M}\neq 0$. On $\cal N$ 
the 1-form $M$ and its complex conjugate $\bar{M}$ may always be augmented by 
a real-valued 1-form $K$ such that 
$M\dz \bar{M}\dz K\neq 0$. Since $\cal N$ is oriented we demand that
$({\rm Re}M, {\rm Im}M, K)$ form a coframe on $\cal N$ which agrees
with the orientation. This coframe is given up to the following 
transformations (null rotations on $\cal N$):
\be
M\to {\rm e}^{i\phi}~M,\quad\quad\quad K\to 
A~[~K+z\bar{M}+\bar{z}M~],\label{eq:transf}
\ee 
where $A>0$, $\phi$ (real) and $z$ (complex) are functions on $\cal N$. The $G$-structure we shall consider corresponds to this group, $G$, of null 
rotations. This 
leads to the following reformulation of the definition for NHSs:
\bd
A null hypersurface structure is a 3-dimensional manifold $\cal N$ equipped with an 
equivalence class of 1-forms $[(M, K)]$ such that 
\begin{itemize}
\item[] $M$ is complex- and $K$ is real-valued,
\item[] $M\dz\bar{M}\dz K\neq 0$ at every point of $\cal N$,
\item[] two pairs $(M,K)$ and $(M_1,K_1)$ are equivalent if and only if  
they are related by transformations (\ref{eq:transf}).
\end{itemize} \label{def:nul2}  
\ed
\bd
Two null hypersurface structures 
$({\cal N},[(M,K)])$ and $({\cal N}_1,[(M_1,K_1)])$ are 
(locally) equivalent iff there exists 
a (local) diffeomorphism $\varphi :{\cal N}\to{\cal N}_1$ and a function 
$\phi:{\cal N}\to [0,2\pi[$ such that $\varphi ^* M_1={\rm e}^{i\phi}~M$.\\ If 
there exists a (local) diffeomorphism $\varphi :{\cal N}\to{\cal N}$ such 
that $\varphi ^* M={\rm e}^{i\phi}~M$ where $\phi$ is a real function, 
then such a $\varphi$ is called a (local) symmetry of a null hypersurface 
structure $({\cal N},[(M,K)])$.\\ An infinitesimal symmetry of 
$({\cal N},[(M,K)])$ 
is a real vector field $X$ on $\cal N$ such that 
$
{\cal L}_X M=i t M,
$  
where $t$ is a real function on $\cal N$.
\ed

\noindent
Let $({\cal N},[(M, K)])$ be a null hypersurface structure as defined above. Let 
$(M,K)$ be a particular representative of the class $[(M, K)])$. Given 
a pair $(M,K)$ we have its differentials which, when decomposed onto
the basis of 2-forms spanned by $M\dz \bar{M}$, $M\dz K$ and
$\bar{M}\dz K$, can be written as follows: 
\beq
&\der M=(\al-\bar{\bet})M\dz \bar{M}+(\eps-\bar{\eps}+\rho)M\dz K+
\bar{\sigma}\bar{M}\dz K,\label{7}\\
&\der K=(\bar{\mu}-\mu)M\dz\bar{M}+(\bar{\pi}-\bet-\bar{\al})M\dz K+
(\pi-\bar{\bet}-\al)\bar{M}\dz\label{8} K.
\eeq
Here $\al,\bet,\eps,\mu,\pi$ are complex functions and $\rho$ is a 
real function on $\cal N$.\\
To describe the above differentials one needs fewer functions than we
have introduced but we write the equations this way so that they agree notationally with equations (\ref{1car}) of Section 2.\\ 
We recall that any null hypersurface $\cal N$ in a space-time $({\cal
M}, g)$ has its natural null hypersurface structure $({\cal
N},h=g_{|\cal N})$. If one has a NHS which originates
from a null hypersurface $\cal N$ in $({\cal M},g)$ then one can locally
introduce a null coframe $(M,\bar{M},L,K)$ on $\cal M$ such that the metric $g$ can be
written as 
$$g=2(M\bar{M}-LK),$$
and the null hypersurface $\cal N$ can be locally defined as a
3-dimensional surface in $\cal M$ such that $L_{|\cal N}\equiv
0$. Such a null tetrad satisfies the equations of Section 2. Now, if
one 
restricts these equations to the null hypersurface $\cal
N$ on which $L\equiv 0$, then the second of these equations imply that
$\rho-\bar{\rho}\equiv\kappa\equiv 0$ on $\cal N$. The first and the
third of these equations coincide with equations
(\ref{7})-(\ref{8}).\\
Thus, although the notation in equations (\ref{7})-(\ref{8}) is
redundant, it has the advantage that the functions appearing in them
can be interpreted as the standard Newman-Penrose coefficients
restricted to a null hypersurface in the case where a NHS originates
from a null hypersurface in a space-time. \\
 Hence we use the notation of equations (\ref{och}) to write, on $\cal
N$, 
\beq
&\G_{41}=\sigma M+\rho\bar{M},\nonumber\\
&\G_{23}=\mu M + \lambda \bar{M}+\pi K,\label{eq:cs1}\\
&\frac{1}{2}(\G_{12}+\G_{34})=-\beta M-\alpha \bar{M}-\epsilon K.\nonumber
\eeq  
\noindent
The functional coefficients in equations (\ref{7})-(\ref{8}) depend on the
choice of a representative $(M,K)$ from the class $[(M,K)]$. In particular we have: 
\bp
Under the gauge transformations (\ref{eq:transf}) the coefficients
$\rho$ and $\sigma$ in equations (\ref{7})-(\ref{8}) transform according to
$$\sigma\to A^{-1}{\rm
e}^{-2i\phi}\sigma,\quad\quad\quad\quad\rho\to A^{-1}\rho.$$ 
\ep
It follows from this proposition that although $\rho$ and $\sigma$
are not well defined objects for a given NHS, their vanishing or not is an
invariant property characterizing a NHS. Thus,  it is clear that NHSs split 
into four
disjoint classes which can not be transformed into each other by
diffeomorphisms. These classes are characterized by 
\begin{itemize}
\item[1)] $\sigma =\rho =0$ on $\cal N$
\item[2)] $\rho =0$ and $\sigma\neq 0$ on $\cal N$
\item[3)] $\sigma\rho\neq 0$ on $\cal N$
\item[4)] $\sigma =0$ and $\rho\neq 0$ on $\cal N$
\end{itemize}
We shall now apply Cartan's method to study the local equivalence
problem, considering each of these cases in turn. The order in which
we consider these inequivalent classes is determined by the way
Cartan's method applies to each of them.
\section{Cartan's invariants for the I - geometry of a null hypersurface}  

In this section we apply Cartan's method and solve the local
equivalence problem for null hypersurface structures $({\cal N},
[(M,K)])$ as defined in the previous section. A particular
representative of $[(M,K)]$, say $(M,K)$, satisfies equations 
(\ref{7})-(\ref{8}) and the equations implied by them.\\

\noindent
To construct the Cartan 
invariants of a NHS we shall need to consider $G$-bundles or
sub-bundles of $G$-bundles over $\cal N$. The highest dimensional such 
bundle will
be ${\cal N}\times\cal F$, where we shall denote the fibre coordinates
of $\cal F$ by the real positive $A$, the complex $z$ and the real
$\phi\in [0,2\pi[$. On occasion subbundles of ${\cal N}\times\cal F$
with a subset of fibre coordinates will be used. Since prolongations
involve replacing gauge group functions by coordinates on $G$-bundles
over $\cal N$ we shall follow the standard shorthand of using the same
symbols $A$, $z$ and $\phi$ for functions on $\cal N$ and coordinates
on the bundles over $\cal N$. The meaning of the symbol will be clear
from the context. On ${\cal N}\times\cal F$ we will have the lifted
coframe\footnote{We use the terminology of 
\cite{bi:Olver}.}, denoted with primes, 
\be
M'={\rm e}^{i\phi}~M,\quad\quad\quad K'=A~[~K+z\bar{M}+\bar{z}M~]
\label{eq:lc}
\ee 
(and analogously on sub-bundles).
Now the usual Cartan procedure of absorbtion and normalization 
(\cite{bi:Olver} pp. 307-309) 
is applied to get the Cartan invariants of the hypersurface. \\
First, we note that the differential of $M'$ can be written in the form  
\be
\der M'=i \omega^1\dz M'+\frac{{\rm e}^{2i\phi}}{A}\bar{\sigma}\bar{M}'\dz K'+
\frac{\rho}{A} M'\dz K'\label{eq:c1}.
\ee
The 1-form $\omega^1$ is not unique. It is 
fixed by requiring that both $\omega^1$ and the coefficient at the 
$M'\dz K'$ term are real. Then, $\omega^1$ reads:
\be
\omega^1=\der\phi+i\G_{21}+i(\bar{z}\bar{\sigma}-z\rho)\bar{M}+i(\bar{z}\bar{\rho}-z\sigma)M.
\label{eq:dm'}
\ee
Equation (\ref{eq:c1}) implies that the Cartan procedure for getting the 
invariants branches according as $\sigma\rho$ is zero or not. 

\subsection{Class 1: $\sigma=\rho=0$ on $\cal N$}
This is a physically interesting case including as it does the
wave-fronts of plane fronted waves, null infinity in the
zero-divergence conformal gauge and horizons, including the recently
introduced `isolated horizons' \cite{bi:Ashtekar}.\\
In this case equation (\ref{eq:c1}) becomes  
\be
\der M'=i\omega^1\dz M'\label{eq:ct1}
\ee
with
\be
\omega^1=\der\phi+i\G_{21}.
\ee 
Calculation of the differential of $K'$ in this case yields
\be
\der K'=\omega^2\dz K'+\bar{\omega}^3\dz M'+\omega^3\dz \bar{M}',
\label{eq:ct2}
\ee
where
\beq
&\omega^2=\der\log A+\G_{34}+bK+cM+\bar{c}\bar{M}\\
&\omega^3=A{\rm e}^{i\phi}~[~\der z +\G_{32}-z(\G_{12}+\G_{34})+(h+\frac{\bar{c}\bar{z}-cz}{2})M+f\bar{M}+(\bar{c}-bz)K~],
\eeq
where we have introduced unknown functions $b,h$ (real) and $f,c$ (complex).\\
Calculations using equations (\ref{eq:ct1}),(\ref{eq:ct2}) show that the original system $({\cal N},[(M,K)])$ does not pass the Cartan test (cf. 
\cite{bi:Olver}, pp. 350-355). Hence it must be prolonged.\\ 
The simplest method of 
prolongation is to first consider a new manifold 
${\cal N}\times{\bf S}^1$ parametrized by 
$(p,\phi)$, where $p\in\cal N$ and $\phi\in [0,2\pi[$. On this manifold one 
considers the system 
of 1-forms $(M',K,\omega^1)$ such that 
\begin{itemize}
\item[i)] $M'$ is complex-, $K$ and $\omega^1$ are real-valued
\item[ii)] $M'\dz\bar{M}'\dz K\dz \omega^1\neq 0$ on ${\cal
N}\times{\bf S}^1$
\item[iii)] $M'={\rm e}^{i\phi} M$, $\omega^1=\der\phi +i\G_{21}$, where the forms $M$, 
$K$ and $\G_{ij}=-\G_{ji}$ satisfy (\ref{eq:cs1}) with $\G_{41}=0$ on 
$\cal N$ and are the lifts to
${\cal N}\times{\bf S}^1$ of the corresponding forms on $\cal N$.
\item[iv)] forms $(M',K,\omega^1)$ are given up to the following
transformations
$$M'\to M',\quad\quad\quad K\to A~[~K+z{\rm
e}^{i\phi}\bar{M}'+\bar{z}{\rm e}^{-i\phi}M'~],\quad\quad\quad \omega^1\to\omega^1,$$ 
where $A>0$ and $z$ are real- and complex functions on 
${\cal N}\times{\bf S}^1$.
\end{itemize}
Now we have a new equivalence problem on ${\cal N}\times{\bf S}^1$
with group $G_1$ given by the transformations in iv). We now prolong
again. Consider now the manifold 
$\tilde{\cal N}={\cal N}\times{\bf S}^1\times{\bf R}_+\times{\bf C}$ 
parametrized by $(q,A,z)$, where 
$q\in{\cal N}\times {\bf S}^1$. The $G_1$-structure, 
when extended to $\tilde{\cal N}$, has the lifted coframe 
$(M',\bar{M}',K',\omega^1)$ with the following differentials:
\beq
&\der M'=i\omega^1\dz M',\nonumber\\
&\der K'=\omega^2\dz K'+\bar{\omega}^3\dz M'+\omega^3\dz \bar{M}',
\label{eq:sys1}\\
&\der\omega^1=i~[-\bar{\Psi}_2-\Psi_2+\frac{S_{12}+S_{34}}{2}+\frac{R}{12}~]~M'\dz \bar{M}',\nonumber
\eeq
where
\beq
&\omega^2=\der\log A+\G_{34}+bK+cM+\bar{c}\bar{M},\\
&\omega^3=A{\rm e}^{i\phi}~[~\der z +\G_{32}-z(\G_{12}+\G_{34})+
(h+\frac{\bar{c}\bar{z}-cz}{2})M+f\bar{M}+(\bar{c}-bz)K~].\nonumber
\eeq
In these formulae $b,h$ (real) and $c,f$ (complex) are undetermined 
functions on ${\cal N}\times{\bf S}^1$, $\Psi_\mu$, $S_{ij}$ and $R$ are the Weyl tensor 
coefficients, the traceless Ricci tensor and the Ricci scalar restricted to 
the null hypersurface and lifted to ${\cal N}\times{\bf S}^1$ 
(by the usual demand that they 
be constant along the $\phi$ direction).\\
The system (\ref{eq:sys1}) has no essential torsions. It passes the 
Cartan test with the following values of the Cartan parameters:
\begin{itemize}
\item[i)] dimension of ${\cal N}\times{\bf S}^1$: n=4,
\item[ii)] dimension of the structure group: $r=3$
\item[iii)] degree of indeterminacy: $r^{(1)}=6$,
\item[iv)] Cartan characters: $s_1'=1$, $s_2'=1$, $s_3'=1$, $s_4'=0$.
\end{itemize}  
Now, assuming that our lifted coframe $(M',\bar{M}',K',\omega^1)$ is 
regular and analytic we can apply Cartan's theorem 
(Theorem 11.16 of Ref. \cite{bi:Olver}) to decide when 
two given NHS's with vanishing relative 
invariants $\sigma$ and $\rho$ are locally equivalent.\\
First we observe, from equations (\ref{eq:sys1}), that the only invariant of a hypersurface with 
$\sigma=\rho=0$ is 
\be
I=-\bar{\Psi}_2-\Psi_2+\frac{S_{12}+S_{34}}{2}+\frac{R}{12}
\ee
and the 
invariants derived from it by differentiation. A simple geometric 
interpretation of $I$ is as follows. Consider a basis 
$(m', \bar{m}',k',o)$ dual to $(M',\bar{M}',K',\omega^1)$ on 
${\cal N}\times{\bf S}^1$. The first equation in the system (\ref{eq:sys1}) guarantees 
that ${\cal N}\times{\bf S}^1$ is foliated by two-dimensional fibres tangent to vector 
fields $k'$ and $o$. Moreover, a $(+ + 0~0)$-signature metric 
$\hat{h}=2M'\bar{M}'$ on ${\cal N}\times{\bf S}^1$ has vanishing Lie 
derivative along the  
directions of $k'$ and $o$. Thus, the metric $\hat{h}$ projects down to a 
well defined metric $h$ on a 2-dimensional manifold 
${\cal N}\times{\bf S}^1/\sim$ of 
the leaves of the foliation\footnote{In other words, any 2-dimensional 
surface $\cal S$ transversal to the leaves of the foliation acquires a 
Riemannian metric $\hat{h}_{|\cal S}$ by restricting $\hat{h}$ to $\cal S$. 
The vanishing of the Lie derivatives means 
that 
for any $\cal S$ the metrics $\hat{h}_{|\cal S}$ are locally isometric.}. 
It can be easily checked that $I$ is the Gaussian 
curvature of $h$.\\

\noindent
The theorem of Cartan (Thm. 11.16 of Ref. \cite{bi:Olver}) implies
that, 
in the fully regular, analytic case there always exist a choice of 
$A$ and $z$ such that $\der K'=0$ in equations 
(\ref{eq:sys1}). Such a choice of $K'$ allows for a
restriction\footnote{By a `restriction' here we mean such a choice of
a section of the bundle ${\cal N}\times{\bf S}^1\to\tilde{\cal N}$ on
which $\der K'=0$ holds. This section is then identified with 
${\cal N}\times{\bf S}^1$.} of 
$\tilde{\cal N}$ to ${\cal N}\times{\bf S}^1$ on which the system 
(\ref{eq:sys1}) takes 
the form 
\beq
&\der M'=i\omega^1\dz M',\nonumber\\
&\der K'=0,\label{eq:sys2}\\
&\der\omega^1=i I M'\dz \bar{M}'.\nonumber
\eeq
It is easy to integrate these equations. The result is: 
\beq
&M'=\frac{{\rm e}^{i\phi}}{P}\der\xi,\nonumber\\
&K'=\der r,\nonumber\\
&\omega^1=\der\phi+i~
[~(\log P)_{\bar{\xi}}\der\bar{\xi}-(\log P)_\xi\der\xi~],\nonumber\\
&I=\triangle\log P,\quad\quad\quad
\triangle=2P^2\partial_{\xi}\partial_{\bar{\xi}},\nonumber
\eeq
where $(\xi,\bar{\xi},r,\phi)$ constitutes a coordinate system on 
${\cal N}\times{\bf S}^1$, $P=P(\xi,\bar{\xi})$ is a real function, and subscripts 
such as $_\xi$ denote partial derivatives. Thus we have the following proposition. 
\bp
Any analytic NHS $({\cal N}, [(M,K)])$ with vanishing relative invariants 
$\sigma$ and $\rho$ is locally equivalent to the one defined by forms 
$$M=\frac{1}{P}{\rm d}\xi,\quad\quad\quad K={\rm d}r,$$
where $P=P(\xi,\bar{\xi})$ is a real function. The lowest order Cartan 
invariant for such NHS's is 
$$I=\triangle\log P,\quad\quad\quad
\triangle=2P^2\partial_{\xi}\partial_{\bar{\xi}}.$$
$I=2c={\rm const}$ if and only if the function $P=1+c \xi\bar{\xi}$.
If $I\neq{\rm const}$, then the higher order invariants are
\begin{itemize}
\item[i)] first order: $I_1=P~I_\xi~{\rm e}^{-i\phi}$
\item[ii)] second order: $I_2=P~(~PI_{\xi\xi}+2P_\xi I_\xi~)~{\rm e}^{-2i\phi}$ and 
$I_3=\frac{1}{2}\triangle I=\frac{1}{2}\triangle\triangle\log P$,
\end{itemize}
where the $\phi$ variable is related to the fourth dimension of the prolonged 
manifold ${\cal N}\times{\bf S}^1$.\\
The local group of symmetries for each such structure is infinite 
dimensional; the set of analytic self-equivalences depends on one real 
function of three real variables. 
\label{pr:00}
\ep
\subsection{Class 2: $\sigma=0$, $\rho\neq 0$ on $\cal N$}
In this case equations (\ref{eq:c1})-(\ref{eq:dm'}) become, on ${\cal
N}\times\cal F$, 
\be
\der M'=i\omega^1\dz M'+\frac{\rho}{A}M'\dz K',
\ee
with 
\be
\omega^1=\der\phi+i\G_{21}-iz\rho\bar{M}+i\bar{z}\bar{\rho}M.\label{no}
\ee
Thus, we can always normalize the coefficient preceeding $M'\dz K'$ to 
$s=\pm 1={\rm sign}(\rho)$ by choosing a section of ${\cal
N}\times{\bf S}^1\times{\bf C}\to{\cal N}\times{\cal F}$ so that
\be
A=|\rho|.
\ee
Using this condition we obtain on ${\cal
N}\times{\bf S}^1\times{\bf C}$
\beq
&\der M'=i\omega^1\dz M'+s M'\dz K'\label{sys2},\\
&\der K'=\bar{\omega}^2\dz M' +\omega^2\dz\bar{M}'\label{sys22},
\eeq 
where $\omega^1$ is given by (\ref{no}), 
\be
\omega^2=s{\rm e}^{i\phi}~[~\rho\der z+(z(\Phi_{01}-\Psi_1)+\rho(2z\bet-\mu))M+(z\Phi_{00}+\bar{\Psi}_1-\bar{\Phi}_{01}+\rho(2z\eps-\pi))K+aM+b\bar{M}~]
\ee
and $a$ is any real function and $b$ is any complex function on ${\cal
N}\times{\bf S}^1\times{\bf C}$. It turns out that the system 
(\ref{sys2})-(\ref{sys22}) does not pass Cartan's 
test. Therefore following the standard procedure we consider the new
equivalence problem for the sub-group preserving the normalisation
above. Hence we consider a new prolongation from $\cal N$ to ${\cal
N}\times{\bf S}^1$ on which we have 
\beq
&\der M'=i\omega^1\dz M'+s M'\dz K'\label{sys3},\\
&\der K'=\bar{\omega}^2\dz M' +\omega^2\dz\bar{M}'\label{sys32},\\
&\der\omega^1=s i\bar{\omega}^2\dz M'-s i\omega^2\dz\bar{M}'+i(
-\bar{\Psi}_2-\Psi_2+\frac{S_{12}+S_{34}}{2}+\frac{R}{12}+2a)M'\dz\bar{M}'.
\nonumber
\eeq
The last of the above equations can be reduced to the form
\be
\der\omega^1=s i\bar{\omega}^2\dz M'-s i\omega^2\dz\bar{M}'\label{sys33}
\ee
by choosing $a=\frac{1}{2}(\bar{\Psi}_2+\Psi_2-\frac{S_{12}+S_{34}}{2}-\frac{R}{12})$. Now, it can be easily checked that 
the system given by equations (\ref{sys3})-(\ref{sys33}) on ${\cal
N}\times{\bf S}^1$ is in involution 
(The Cartan parameters are: $n=4$, $r=2$, $r^{(1)}=2$, $s_1'=2$, $s_2'=0$, 
$s_3'=0$, $s_4'=0$), so 
in the analytic fully regular case, it is gauge equivalent to the system 
\beq
&\der M'=i\omega^1\dz M'+sM'\dz K'\nonumber\\ 
&\der K'=0\label{sys4}\\
&\der\omega^1=0.\nonumber
\eeq
This means that in this case where $\sigma=0$, $\rho\neq 0$ we have only 
two locally non-equivalent null hypersurfaces. They correspond to the
choices $s=+1$ and $s=-1$ in equations (\ref{sys4}).\\ 

\noindent
The system of equations (\ref{sys4}) can be easily solved. The result is:
\beq
&M'={\rm e}^{i\phi\mp r}\der\xi\\
&K'=\der r\\
&\omega^1=\der \phi,
\eeq
where $(\xi,\bar{\xi},r,\phi)$ constitutes a coordinate system on ${\cal
N}\times{\bf S}^1$.
\\In this way we obtained the following proposition.
\bp
Any analytic null hypersurface structure $({\cal N},[(M,K)])$ with 
vanishing relative invariant 
$\sigma$ and nonvanishing $\rho$ is locally equivalent to one of the
two 
NHS's defined by the forms
\beq
&M={\rm e}^{- r}{\rm d}\xi\\
&K={\rm d} r
\eeq
or by the forms 
\beq
&M={\rm e}^r{\rm d}\xi\\
&K={\rm d} r
\eeq
on a manifold $\cal N$ with coordinates $(\xi,\bar{\xi},r)$.\\
The local group of symmetries for such structures is infinite dimensional; the set of analytic self-equivalences depends on two real functions of one real 
variable.
\ep
We note that past and the future null cones in Minkowski space-time 
possess the NHS's of Proposition 3.\\

\noindent
In dealing with the last two classes we continue to use methods and
notation similar to those used above. However, as the essential
features of Cartan's method are algorithmic and repetitive we
increasingly abbreviate. The reader interested only in the final
results of the calculations will find them stated as propositions at
the ends of the sections.
\subsection{Class 3: $\sigma\neq 0$, $\rho\neq 0$}
In this, the generic case, we can always normalize the factor of 
the $\bar{M}'\dz K'$ term of (\ref{eq:c1}) to 1. We achieve this by 
imposing condition
$A{\rm e}^{2i\phi}=\sigma$ on the group parameters $A,\phi$. One sees that 
this condition fixes both $A$ and $\phi$. From now, on we assume that a 
representative $(M,K)$ defining a null surface structure on $\cal N$
has been gauged to the form in which equation (\ref{7}) is 
\be
\der M=(\al-\bar{\bet})M\dz \bar{M}+(\eps-\bar{\eps}+\rho)\bar{M}\dz K+
\bar{M}\dz K.
\ee
This condition is equivalent to the statement that $(M,K)$ has been 
gauged to the form in which 
\be
\sigma=1.\label{s1}
\ee 
After this has been done, 
the remaining gauge freedom is 
$$M\to s M,\quad\quad\quad K\to K+z\bar{M}+\bar{z}M,$$
where $s=\pm 1$.
Thus we have a new equivalence problem defined by the above 
transformations. To construct the Cartan invariants we consider a manifold 
${\cal N}\times{\bf C}$ parametrized by $(p,z)$, 
$p\in\cal N$, $z\in{\bf C}$. Then a straightforward calculation, using
as before lifts of $M$ and $K$, shows that 
the differential of $M'$ is
\be
\der M'=s ~[~\bar{z}-z(\rho+\eps-\bar{\eps})+\al-\bar{\bet}~]~M'\dz\bar{M}'+
(\rho+\eps-\bar{\eps})M'\dz K'+\bar{M}'\dz K'.\label{e5}
\ee
This equation implies that the function  
\be
I^0_1=\rho+\eps-\bar{\eps}
\ee
constitutes an invariant of the NHS\footnote{Remember that this 
expression for the invariant $I^0_1$ 
is valid only in the gauge (\ref{s1}).}.\\

\noindent
Now, two cases may occur. Either
$$(i)\quad\quad
|I^0_1|\neq 1\quad\quad\quad{\rm (the~generic~ case)}
$$
or 
$$(ii)\quad\quad
|I^0_1|=1\quad\quad\quad{\rm (the~special~ case)}.
$$

\noindent
{\bf Case (i): $|I^0_1|\neq 1$.}\\

\noindent
If $|I^0_1|\neq 1$ we choose
\be
z=\frac{\bar{I}^0_1(\al-\bar{\bet})+\bar{\al}-\bet}{|I^0_1|^2-1}\label{ez}
\ee
to reduce equation (\ref{e5}) to the form (on $\cal N$)
\be
\der M'=I^0_1 M'\dz K'+\bar{M}'\dz K'.\label{5}
\ee
\noindent
Now, the only remaining freedom is in the choice of a sign $s$. To fix
this, 
we remark that
equation (\ref{5}) implies that a gauge can be chosen in which the
initial differential forms $(M,K)$ satisfy equations (\ref{7}), (\ref{8}) with 
$\sigma=1$ and $\al=\bar{\bet}$. In such a gauge the differential of
the form $K'$ is (on $\cal N$)
\be
\der K'= i I^0_2M'\dz \bar{M}'+I^0_3M'\dz K'+\bar{I}^0_3\bar{M}'\dz K',
\label{51}
\ee 
where the coefficients $I^0_i$ are:
\be
I^0_1=\rho+\eps-\bar{\eps},\quad\quad\quad I^0_2=i(\mu-\bar{\mu}),
\quad\quad\quad I^0_3=s(\bar{\pi}-2\bar{\al}).
\ee 
(Note, that since $\rho\neq 0$, $I^0_1+\bar{I}^0_1\neq 0$).\\ 
Now, assuming that $I^0_3\neq 0$ we can fix $s$ in such a way that $s{\rm Re} [I^0_3]>0$ 
(if ${\rm Re} [I^0_3]\neq 0$) or 
$s{\rm Im} [I^0_3]>0$ (if ${\rm Re} [I^0_3]=0$). If $I^0_3=0$ there is
no way to fix the sign $s$. 

\noindent
We summarize the result of this section in the following proposition.

\bp
The generic NHS $({\cal N}, [(M,K)])$
($\sigma\neq 0,~\rho\neq 0,~|I^0_1|\neq 1,~I^0_1+\bar{I}^0_1\neq 0$) 
defines invariant forms 
$(M,K)$ which satisfy 
\beq
&{\rm d} M=I^0_1 M\dz K+\bar{M}\dz K\label{61}\\
&{\rm d} K= i I^0_2M\dz \bar{M}+I^0_3M\dz K+\bar{I}^0_3\bar{M}\dz K,
\label{62}
\eeq
in a unique way if $I^0_3\neq 0$, and with a sign ambiguity in the choice of $M$ if $I^0_3= 0$. The scalar 
invariants of zeroth order for such a NHS are $I^0_1$ (such that 
$|I^0_1|\neq 1$), $I^0_2$ (real) and $I^0_3$ (in general complex). All the 
other invariants are derived from those by means of differentiation.\\
Any such null hypersurface has a local group of symmetries of
dimension no greater than 3.
\ep

\noindent
The system (\ref{61})-(\ref{62}) is too general to be as easily
integrated as was the case in the systems studied in preceding sections. 
Here we only consider the system (\ref{61})-(\ref{62}) 
when it has a 3-dimensional group of symmetries acting transitively\footnote{Hence the classification of inequivalent generic zeroth
order NHS's is in correspondence with the 
Bianchi classification of 3-dimensional Lie algebras} on 
$\cal N$, that is, all the 
invariants $I^0_1$, $I^0_2$ and $I^0_3$ are constants on $\cal N$ - 
zeroth order NHS's. Then the equations ${\rm d}^2M=0$, ${\rm d}^2K=0$,
consequences of equations (\ref{61})-(\ref{62}), 
imply that $I^0_2=I^0_3=0$. The other
possibility $I^0_1+\bar{I}^0_1=I^0_3=0$ is excluded by the 
assumption $\rho\neq 0$.

\bp
Any NHS $({\cal N}, [(M,K)])$ with $I^0_2=I^0_3=0$ and 
$I^0_1=a+i b$, 
$a,b$={\rm const}$\in {\bf R}$ is equivalent to the one defined by
forms on $\cal N$ given by  
$$M=[\bar{\xi}+I^0_1\xi]{\rm d}r+{\rm d}\xi,\quad\quad\quad K={\rm d}r.$$
The three dimensional algebra of infinitesimal symmetries of such a 
NHS  is of Bianchi type\\
i) $IV$ iff $b=\pm 1$,\\
ii) $VI_{-h}$, $h=\frac{a^2}{b^2-1}$ iff $|b|<1$,\\
iii) $VII_h$, $h=\frac{a^2}{b^2-1}$ iff $|b|>1$.\\    
\ep
Thus, there are only two nonequivalent generic NHS's of zeroth order 
admitting the symmetry algebra of Bianchi type $IV$. For each value of 
the parameter $h$, there is a one-parameter 
family of nonequivalent NHS's of zeroth order of Bianchi types 
$VI_{-h}$ and $VII_h$.\\

\noindent
{\bf Case (ii): $|I^0_1|=1$.}\\

\noindent
If $|I^0_1|=1$ there exists a real function $t:{\cal N}\to [0,2\pi[$ such that $I^0_1={\rm e}^{it}$. With 
this choice equation (\ref{e5}) assumes the form 
\be
\der M'=s ~[~\bar{z}-z{\rm e}^{it}+\al-\bar{\bet}~]~M'\dz\bar{M}'+ {\rm e}^{it}M'\dz K'+\bar{M}'\dz K'.
\label{e27}
\ee
Now the following two cases may occur
\begin{itemize}
\item[ii.a)] $t\neq\pi$
\item[ii.b)] $t= \pi.$
\end{itemize}

\newpage
\noindent
{\bf Case ii.a) ($t\neq\pi$)}\\

\noindent
Let ${\rm Re} [\al-\bar{\bet}]=A$, ${\rm Im}[\al-\bar{\bet}]=B$, ${\rm Re}[z]=\zeta$, ${\rm Im}[z]=\eta$. 
Since $t\neq\pi$ we can 
choose the function $\eta$ in the form \be\eta=\frac{B-\zeta\sin t}{1+\cos t}.\ee  Such a choice of 
$\eta$ corresponds to a choice of $z$ that makes the coefficient of
the $M'\dz\bar{M}'$ term in equation (\ref{e27}) real.
When this choice has been made we have 
\be
\der M'=saM'\dz\bar{M}'+ {\rm e}^{it}M'\dz K'+\bar{M}'\dz K',\label{e28}
\ee
where $a$ is a real function. Now, the remaining gauge freedom is 
\be M'=sM\quad\quad\quad\quad K'=K+r({\rm e}^{-it/2}\bar{M}+{\rm e}^{it/2}M),
\ee
where $r$ is any real function. It is now convenient to change the variables 
from $(M, K)$ to $(N,K)$, 
where 
\be 
N={\rm e}^{it/2}M.
\ee 
This choice is admissible since $t$ is an invariant. The forms $(N,K)$ are given up to transformations 
\be
N'=sN,\quad\quad s=\pm 1,\quad\quad\quad\quad K'=K+r(N+\bar{N}),\label{e222}
\ee
where $r$ is a real function. Equation (\ref{e28}) rewritten in terms of $(N',K')$ reads
\be
\der N'=\frac{i}{2}\der t\dz N'+sa{\rm e}^{it/2}N'\dz\bar{N}'+{\rm e}^{it}(N'+\bar{N}')\dz K'.\label{eq:222}
\ee
Thus we have a new equivalence problem for the forms $(N,K)$ which are given up to the transformations (\ref{e222}). 
The form $N'$ satisfies (\ref{eq:222}). Writing the differential of $K$ in the most general form
\be
\der K=i bN\dz\bar{N}+cN\dz K+\bar{c}\bar{N}\dz K,
\ee
and introducing the notation $\der t=t_kK+t_+(N+\bar{N})+t_-i(N-\bar{N})$ we easily find that
\be
\der K'=s\omega^0\dz(N'+\bar{N}')+s[\frac{1}{2}(\bar{c}-c)+\frac{i}{2}rt_k]K'\dz (N'-\bar{N}'),\label{pip}
\ee
where 
\be
\omega^0=\der r-[2r\cos t+\frac{1}{2}(c+\bar{c})]K+\frac{1}{2}[i(b-r t_++2ra\sin\frac{t}{2})+r(\bar{c}-c+irt_k)]
(N-\bar{N})+e(N+\bar{N}),
\ee
and $e$ is an arbitrary real function on $\cal N$.\\
Now the following two cases may occur
\begin{itemize}
\item[ii.a1)] $t_k\neq 0$ (i.e. $\der t\dz N\dz\bar{N}\neq 0$)
\item[ii.a2)] $t_k=0$ (i.e. $\der t\dz N\dz\bar{N}= 0$.)
\end{itemize}

\noindent 
{\bf Case ii.a1) ($\der t\dz N\dz\bar{N}\neq 0$)}\\

\noindent
If $\der t\dz N\dz\bar{N}\neq 0$ then the choice 
\be
r=\frac{c-\bar{c}}{it_k}
\ee
brings the system given by equations (\ref{eq:222}), (\ref{pip}) to the form 
\beq
&\der N'=\frac{i}{2}\der t\dz N'+sa{\rm e}^{it/2}N'\dz\bar{N}'+{\rm e}^{it/2}(N'+\bar{N}')\dz K'\\
&\der K'=ib'N'\dz\bar{N}'+c'(N'+\bar{N}')\dz K',
\eeq
where $b'$ and $c'$ are appropriate functions on $\cal N$.\\
Now, we can return to the original variables $(M,K)$ instead of $(N,K)$. The result is summarized 
in the following proposition.
\bp
A generic NHS $({\cal N}, [(M,K)])$ for which $I^0_1={\rm e}^{it}$, is equivalent to the 
one generated by forms 
$(M,K)$ which satisfy 
\beq
&\der M=saM\dz \bar{M}+{\rm e}^{it}M\dz K+\bar{M}\dz K\\
&\der K=isbM\dz\bar{M}+sc[{\rm e}^{it/2}M+{\rm e}^{-it/2}\bar{M}]\dz K.\label{bu}
\eeq
Here $\der t\dz M\dz\bar{M}\neq 0$, $a\geq 0$, $b$ and $c$ are real 
functions on $\cal N$. 
The full system of invariants in this case is given by $t,~sa,~sb,~sc$ and their differentials. The local group of symmetries 
of such structures is at most of dimension 2.
\ep
(We note that the word ``generic''in the above definition means 
``$\der t\dz M\dz\bar{M}\neq 0$'').\\

\noindent
We close this section remarking that the forms $(M,K)$ are determined by conditions (\ref{bu}) uniquely 
modulo the sign of $M$. If not all the invariants $a,~b,~c$ are 
simultaneously zero, one can fix this freedom 
by requiring that $sa$ or $sb$ or $s{\rm Re}c$ or $s{\rm Im}c$ is greater than zero.\\

\noindent
{\bf Case ii.a2) ($\der t\dz N\dz\bar{N}= 0$).}\\

\noindent
In this case equations (\ref{eq:222}), (\ref{pip}) take the form 
\beq
&\der N'=\frac{i}{2}\der t\dz N'+sa{\rm e}^{it/2}N'\dz\bar{N}'+{\rm e}^{it}(N'+\bar{N}')\dz K',\nonumber\\
&\der K'=s\omega^0\dz(N'+\bar{N}')+s\frac{1}{2}[\bar{c}-c]K'\dz (N'-\bar{N}').\label{nn}
\eeq
The system (\ref{nn}) passes the Cartan test (with $n=3$, $r=1$, $r^{(1)}=1$, $s_1'=1$, $s_2'=0$, and $s_3'=0$). 
Thus we conclude that it is gauge equivalent to the system 
\beq
&\der N'=\frac{i}{2}\der t\dz N'+sa{\rm e}^{it/2}N'\dz\bar{N}'+{\rm e}^{it}(N'+\bar{N}')\dz K',\nonumber\\
&\der K'=s\frac{1}{2}[\bar{c}-c]K'\dz (N'-\bar{N}').
\eeq
Returning to the original variables $(M,K)$ we have the following proposition.
\bp
A NHS $({\cal N}, [(M,K)])$ for which $I^0_1={\rm e}^{it}$, $\der I^0_1\dz M\dz\bar{M}=0$, $I^0_1\neq -1$ is equivalent 
to the one generated by the forms $(M,K)$ satisfying the following equations
\beq
&\der M=sa M\dz\bar{M}+{\rm e}^{it}M\dz K+\bar{M}\dz K\nonumber\\
&\der K=isb[{\rm e}^{it/2}M\dz K-{\rm e}^{-it/2}\bar{M}\dz K],
\eeq 
where $a$ and $b$ are real functions, $I^0_1={\rm e}^{it}$.\\
The invariants of the above structures are $a,~b,~t$ and their differentials. The sign ambiguity in the 
choice of $M$ can be removed by the requirements that $sa$ or $sb$ is greater than zero (provided that both 
$a$ and $b$ do not vanish).\\
The symmetry group of any of such structures is infinitely dimensional. The set of analytic 
self-equivalences for each structure depends on one real function of one 
variable.
\ep

\noindent
As an example of null surfaces described by the above proposition all such structures for which $a$, $b$ and $t$ are constants are given in the following 
result:\\

\noindent
Hypersurfaces for which $|I^0_1|=1$, $\der I^0_1\dz M\dz\bar{M}=0$, $I^0_1\neq -1$, $\der a=\der b=\der t=0$ have neccessarily $a=b=0$. Locally they can be described in a coordinate chart 
$(r,\xi,\bar{\xi})$ by the forms 
\be
M=\der\xi +({\rm e}^{it}\xi+\bar{\xi})\der r,\quad\quad\quad\quad K=\der r,\quad\quad\quad t\neq \pi.
\ee

\noindent
{\bf Case ii.b) ($t=\pi$).}\\

\noindent
If $t=\pi$ then equation (\ref{e27}) takes the form 
\be
\der M'=s ~[~\bar{z}+z+\al-\bar{\bet}~]~M'\dz\bar{M}'-M'\dz K'+\bar{M}'\dz K'.
\ee
We fix Re$z$ by demanding that the coefficient of the $M'\dz \bar{M}'$ term is 
purely imaginary. Hence in this case we start with the forms 
$(M,K)$ such that
\be
\der M=i a M\dz \bar{M}-M\dz K+\bar{M}\dz K,
\ee 
where $a$ is a nonnegative real function.\\
The remaining gauge freedom is 
\be
M'=sM,\quad\quad s=\pm 1,\quad\quad\quad\quad K'=K+ir(M-\bar{M}),
\ee
where $r$ is arbitrary real function.
Writing the differential of $K$ in the most general way
\be
\der K=ib M\dz\bar{M}+cM\dz K+\bar{c}\bar{M}\dz K,
\ee
where $b$ is real function, we get that following equations for the differentials of the 
lifted forms $(M',K')$.
\beq
&\der M'=is a M'\dz \bar{M}'-M'\dz K'+\bar{M}'\dz K',\nonumber\\
&\der K'=is\omega^0\dz (M'-\bar{M}')+\frac{1}{2}s(c+\bar{c})(M'+\bar{M}')\dz K'\\
&\omega^0=\der r+[2r+\frac{i}{2}(c-\bar{c})]K-\frac{1}{2}[b+ir(c++\bar{c})](M+\bar{M})+
iu(M-\bar{M})\nonumber
\eeq 
where $u$ is a real function.\\
The above system passes Cartan's test with $n=3$, $r=1$, $r^{(1)}=1$, $s_1'=1$, $s_2'=s_3'=0$. 
This leads to the following proposition.
\bp
Any NHS $({\cal N}, [(M,K)])$ with $I^0_1=-1$ is gauge equivalent to one of the structures generated by 
the forms $(M,K)$ satisfying 
\beq
&\der M=i a M\dz \bar{M}+K\dz (M-\bar{M}),\nonumber\\
&\der K=b(M+\bar{M})\dz K,
\eeq
where $a$ and $b$ are real functions. If $a=0$ and $b=0$ the forms $(M,K)$ are defined by the 
above equations up to the sign of $M$, otherwise the forms are defined uniquely.\\
The full set of invariants for such NHS's consists of the 
functions $a$, $b$ and their 
differentials. If $a\neq 0$ one 
can restrict to $a>0$. If $a=$ and $b\neq 0$ it is enough to consider $b>0$.\\
The symmetry group of such hypersurfaces is infinite dimensional. The set of all 
analytic self equivalences depends on one real function of one variable.
\ep
As an example of the hypersurfaces described by the above proposition we consider 
the case where $a$ and $b$ are 
constants. Then it follows that they both must be zero, and there exists a coordinate chart 
$(r,\xi,\bar\xi)$ on such $({\cal N},[(M,K)])$ such that 
\be
M=\der \xi+(\xi-\bar{\xi})\der r,\quad\quad\quad K=\der r.
\ee

\subsection{Class 4: $\rho =0$ and $\sigma\neq 0$ on $\cal N$}
This case is very special because the conditions $\kappa=\rho=0$ on 
$\cal N$, together with the second structure Cartan equations (7.28) of 
\cite{bi:Kramer} imply  that 
\be
\sigma\bar{\sigma}+S_{44}/2=0\quad{\rm on}\quad{\cal N}.\label{eq:cc}
\ee
Usually one is interested in the case where $\cal N$ is a hypersurface 
in a 
physically 
realistic space-time. Here we see that any hypersurface on which 
$\rho =0$ and $\sigma\neq 0$ can not exist in a space-time satisfying the dominant energy condition $S_{44}\geq 0$; hence, in particular, such NHS can not exist in any vaccum space-time.\\

\noindent 
If $\rho =0$ and $\sigma\neq 0$ then the coefficient of the $\bar{M}'\dz K'$ term in equation 
(\ref{eq:c1}) may be normalized to 1. Then we have 
\be
\bar{\sigma}A^{-1}{\rm e}^{2i\phi}=1
\ee
which totally fixes $A$ and determines ${\rm e}^{i\phi}$ up to a sign. Hence in this case 
we can always start with forms $(M,K)$ which satisfy 
\be 
\der M=(\al-\bar{\bet})M\dz\bar{M}+(\eps-\bar{\eps})M\dz K+\bar{M}\dz K.
\ee
The remaining gauge freedom is 
\be
M'=sM,\quad\quad\quad s=\pm 1,\quad\quad\quad\quad K'=K+z\bar{M}+\bar{z}M,
\ee
where $z$ is a complex function.\\
We now proceed in the usual way. Calculating the differential of $M'$ we get
\be
\der M'=s[\bar{z}-z(\eps-\bar{\eps})+\al-\bar{\bet}]M'\dz\bar{M}'+
(\eps-\bar{\eps})M'\dz K'+\bar{M}'\dz K'.\label{ziut}
\ee
Thus we see that the quantity $iJ^0_1=\eps-\bar{\eps}$ is an invariant.\\

\noindent
The following two cases are worth distinguishing.
\begin{itemize}
\item[a)] $J^0_1\neq\pm 1$
\item[b)] $J^0_1=\pm 1$.
\end{itemize} 

\noindent
{\bf Case a) $J^0_1\neq\pm 1$.}

\noindent
In this case we can fix $z$ by 
\be
z=\frac{-i\bar{J}^0_1(\al-\bar{\bet})+\bar{\al}-\bet}{|J^0_1|^2-1}.
\ee
Such a choice eliminates the $M'\dz\bar{M}'$ term. 
The remaining freedom is only in the sign of $M$. The system reduces to 
\beq
\der M'=iJ^0_1M'\dz K'+\bar{M}'\dz K'\nonumber\\
\der K'=iJ^0_2M'\dz\bar{M}'+sJ^0_3M'\dz K'+s\bar{J}^0_3\bar{M}'\dz K',
\eeq
where $J^0_2$ is a real function and $J^0_3$ is a complex function.\\
If $J^0_3\neq 0$ then we can fix the freedom totally by the requirement that its real 
or imaginary part is positive. Thus we have the following proposition. 
\bp
Any NHS $({\cal N}, [(M,K)])$ for which $\rho=0,\sigma\neq 0$ and $J^0_1\neq\pm 1$ is equivalent to one of the NHS's defined by the forms $(M,K)$ satisfying 
\beq
\der M=iJ^0_1M\dz K+\bar{M}\dz K\nonumber\\
\der K=iJ^0_2M\dz\bar{M}+J^0_3M\dz K+\bar{J}^0_3\bar{M}\dz K,\label{ziu}
\eeq
where ${\rm Re}J^0_3\geq 0$ (or ${\rm Im}J^0_3\geq 0$ if ${\rm Re}J^0_3= 0$). The full set of 
invariants for such NHS's consists of $J^0_1,~J^0_2,~J^0_3$ and their differentials. 
The forms $(M,K)$ are defined by equation (\ref{ziu}) uniquely if 
$J^0_3\neq 0$ and up to the sign of $M$ if 
$J^0_3=0$.\\
Every such NHS has symmetry group of dimension no greater than 3.
\ep 
The classification of NHS's with $J^0_1\neq\pm 1$ and symmetry group of dimension 3 is given in Appendix A.\\

\noindent
{\bf Case b) $J^0_1=\pm 1$.}\\

\noindent
Since $J^0_1$ is an invariant the null hypersurfaces belonging to the classes characterized by the 
opposite sign of $J^0_1$ are not equivalent. To find invariants in each of the two classes we proceed as 
follows.\\ 
Equation (\ref{ziut}) implies that  
\be
\der M'=s[\bar{z}\mp iz+\al-\bar{\bet}]M'\dz\bar{M}'\pm 
iM'\dz K'+\bar{M}'\dz K'.
\ee
Writing $z$ and $\al-\bar{\bet}$ in the form $z=\zeta+i\eta$ and $\al-\bar{\bet}=A+iB$, with $\zeta,\eta,A,B$ real, 
we see that 
the choice $z=iB+\zeta(1\mp i)$ makes the coefficient of the $M'\dz\bar{M}'$ term real. 
This shows that the present case can be 
generated by starting with forms $(M,K)$ satisfying 
\beq
&\der M=aM\dz\bar{M}\pm i 
M\dz K+\bar{M}\dz K,\nonumber\\
&\der K=ibM\dz\bar{M}+cM\dz K+\bar{c}\bar{M}\dz K,
\eeq 
where $a,b$ (real) and $c$ (in general complex) are functions. The remaining 
gauge freedom is 
\be
M'=sM,\quad\quad\quad s=\pm 1,\quad\quad\quad\quad K'=K+r[(1\mp i)\bar{M}+(1\pm i)M],
\ee
where $r$ is a real function.\\ 
The differentials of $(M',K')$ read
\beq
&\der M'=saM'\dz\bar{M}'\pm 
iM'\dz K'+\bar{M}'\dz K',\nonumber\\
&\der K'=s\omega^0\dz [(1\pm i)M'+(1\mp i)\bar{M}']+\label{taf}\\
&\frac{1}{4}s[c(1\mp i)-\bar{c}(1\pm i)]
[(1\pm i)M'-(1\mp i)\bar{M}']\dz K'\nonumber,
\eeq
where
\beq
&4\omega^0=4\der r+[ib\pm 2iar-cr(1\mp i)+\bar{c}r(1\pm i)][(1\pm i)M-(1\mp i)\bar{M}]\nonumber\\
&-[c(1\mp i)+\bar{c}(1\pm i)]K+
u[(1\pm i)M+(1\mp i)\bar{M}]
\eeq
and $u$ is arbitrary real constant. We easily verify that the system (\ref{taf}) passes the Cartan test 
($n=3$, $r=1$, $r^{(1)}=1$, $s_1'=1$, $s_2'=s_3'=0$), hence we have the following proposition.
\bp
Every NHS $({\cal N}, [(M,K)])$ for which $J^0_1=\pm 1$ is gauge equivalent to one of the null 
hypersurfaces generated by forms $(M,K)$ with differentials 
\beq
&\der M=aM\dz\bar{M}\pm 
iM\dz K+\bar{M}\dz K,\nonumber\\
&\der K=b[(1\pm i)M-(1\mp i)\bar{M}]\dz K\nonumber,
\eeq
where $a\geq 0$, $b$ are real functions ($b\geq 0$ if $a=0$).\\ The full set of invariants for such NHS's is 
given by $a$, $b$ and their differentials.\\ The forms $(M,K)$ are defined uniquely if at least one of $a$ and 
$b$ does not vanish. Otherwise the forms are defined up to a sign of $M$.\\
The symmetry group for such NHS's is infinite dimensional. The set of all analytic 
self-equivalences for each NHS depends on one real function of one variable.
\ep

An example of a surface belonging to the class described by the above proposition is a null hypersurface structure defined by forms 
\be
M=\der\xi+(\bar{\xi}\pm i\xi)\der r,\quad\quad\quad\quad K=\der r,
\ee
on a surface $\cal N$ coordinatized by $(r,\xi,\bar{\xi})$.\\

\noindent
All the major results of this section and those of Appendix A are summarized in a diagram in Appendix B.

\section{Conclusions}
We have investigated the intrinsic geometry of null hypersurfaces by 
formulating the concept of a null hypersurface structure. All such 
structures have been classified into equivalence classes by using Cartan's 
method which is adapted to the use of differential forms. The resulting 
classification can be used in the study of gravitational problems involving 
null hypersurfaces. The techniques that have been used could also be 
applied to null hypersurfaces in higher dimensional space-times. Here we 
have concentrated on classifying a structure which is equivalent to 
Penrose's I - geometry. It would be interesting and informative to 
classify II - and II - geometries and to compare the solutions of the 
corresponding equivalence problems.
\newpage
\appendix
\section{NHS's with 3-dimensional symmetry groups.}
Here we explicitly compute the details of one of the classes specified in Proposition 9 of Section 4. We present all null NHS's for which 
$\rho=0$, $\sigma\neq 0$ and the 
symmetry group is strictly 3-dimensional. They have the property that all the invariants 
$J^0_1,~J^0_2,~J^0_3$ are constants.\\ 
By taking the exterior derivative of equations (\ref{ziu}) we obtain 
$J^0_3=0$. Thus 
such NHS'sare defined by the system (cf. equation (\ref{ziu}))
\beq
\der M=iJ^0_1M\dz K+\bar{M}\dz K\nonumber\\
\der K=iJ^0_2M\dz\bar{M},\nonumber
\eeq
with real constants $J^0_1$ and $J^0_2$. The following cases may occur.
\begin{itemize}
\item[1)] $J^0_2\neq 0$
\begin{itemize}
\item[1a)] $(-J^0_2,J^0_1-1,J^0_1+1)$ have all the same sign $s$.\\
In this case the symmetry group is of Bianchi type $IX$, and the forms $(M,K)$ may be 
choosen in such a way that
\be
M=\frac{s(1+i)}{2\sqrt{-J^0_2(J^0_1-1)}}\theta^1+\frac{1-i}{2\sqrt{-J^0_2(J^0_1+1)}}\theta^2,\quad\quad\quad
K=\frac{1}{\sqrt{(J^0_1)^2-1}}\theta^3,\nonumber
\ee
where the forms $(\theta^1,\theta^2,\theta^3)$ in a suitable coordinate system $(u,x,y)$ are given by
\beq
&\theta^1=\cos y\cos u\der x-\sin u\der y\nonumber\\
&\theta^2=\cos y\sin u\der x+\cos u\der y\nonumber\\
&\theta^3=-\sin y\der x+\der u.\nonumber
\eeq  
We have a 2-parameter family of nonequivalent NHS's here.
\item[1b)] not all $(-J^0_2,J^0_1-1,J^0_1+1)$ have the same sign.\\
In this case the symmetry group is of Bianchi type $VIII$, so it is convenient to introduce 
a coordinate chart $(x,y,u)$ on $\cal N$ and forms  
\beq
&\Theta^1=\cosh y\cos u\der x-\sin u\der y\nonumber\\
&\Theta^2=\cosh y\sin u\der x+\cos u\der y\nonumber\\
&\Theta^3=\sinh y\der x+\der u.\nonumber
\eeq  
The following three cases may occur.
\begin{itemize}
\item[1bi)] $J^0_1+1=s|J^0_1+1|$, $J^0_1-1=s|J^0_1-1|$, $J^0_2=s|J^0_2|$, $s=\pm 1$.\\
In this case the null hypersurface structure is generated by $(M,K)$ such that  
\be
M=\frac{s(1+i)}{2\sqrt{J^0_2(J^0_1-1)}}\Theta^1+\frac{1-i}{2\sqrt{J^0_2(J^0_1+1)}}\Theta^2,\quad\quad\quad
K=\frac{1}{\sqrt{(J^0_1)^2-1}}\Theta^3.\nonumber
\ee
\item[1bii)] $J^0_1+1=s|J^0_1+1|$, $J^0_1-1=-s|J^0_1-1|$, $J^0_2=s|J^0_2|$, $s=\pm 1$.\\
In this case the null hypersurface structure is generated by $(M,K)$ such that  
\be
M=\frac{s(1+i)}{2\sqrt{-J^0_2(J^0_1-1)}}\Theta^3+\frac{1-i}{2\sqrt{J^0_2(J^0_1+1)}}\Theta^2,\quad\quad\quad
K=\frac{1}{\sqrt{1-(J^0_1)^2}}\Theta^1.\nonumber
\ee
\item[1biii)] $J^0_1+1=s|J^0_1+1|$, $J^0_1-1=-s|J^0_1-1|$, $J^0_2=-s|J^0_2|$, $s=\pm 1$.\\
In this case the null hypersurface structure is generated by $(M,K)$ such that  
\be
M=\frac{1+i}{2\sqrt{J^0_2(J^0_1-1)}}\Theta^2+\frac{s(1-i)}{2\sqrt{-J^0_2(J^0_1+1)}}\Theta^3,\quad\quad\quad
K=\frac{1}{\sqrt{1-(J^0_1)^2}}\Theta^1.\nonumber
\ee
\end{itemize}
In all cases 1b) we have 2-parameter families of inequivalent NHS's.
\end{itemize}
\item[2)] $J^0_2= 0$
\begin{itemize}
\item[2a)] $|J^0_1|<1$.\\
In this case the symmetry group is of Bianchi type $VI_0$, and the forms $(M,K)$ may be 
chosen in such a way that
\be
M=\frac{1+i}{\sqrt{2}}\sqrt{\frac{1+J^0_1}{1-J^0_1}}~\tau^1+\frac{1-i}{\sqrt{2}}~\tau^2,\quad\quad\quad
K=\frac{1}{\sqrt{1-(J^0_1)^2}}~\tau^3,\nonumber
\ee
where, in appropriate coordinates, the forms $(\tau^1,\tau^2,\tau^3)$ are
\beq
&\tau^1=\cosh x\der y-\sinh x\der u\nonumber\\
&\tau^2=-\sinh x\der y+\cosh x\der u\nonumber\\
&\tau^3=\der x.\nonumber
\eeq  
It follows that we have a one-parameter family of nonequivalent NHS's here.
\item[2b)] $|J^0_1|>1$\\
In this case the symmetry group is of Bianchi type $VII_0$, and the forms $(M,K)$ may be 
chosen in such a way that
\be
M=\frac{1+i}{\sqrt{2}}\sqrt{\frac{J^0_1+1}{J^0_1-1}}~\si^1+\frac{1-i}{\sqrt{2}}~\si^2,\quad\quad\quad
K=\frac{1}{\sqrt{(J^0_1)^2-1}}~\si^3,\nonumber
\ee
where, in appropriate coordinates, the forms $(\si^1,\si^2,\si^3)$ are
\beq
&\si^1=\cos x\der y-\sin x\der u\nonumber\\
&\si^2=\sin x\der y+\cos x\der u\nonumber\\
&\si^3=\der x.\nonumber
\eeq  
It follows that here we have a one-parameter family of inequivalent null hypersurfaces. 
\end{itemize}
\end{itemize}
We close this appendix by remarking that the NHS's described here and in Proposition 5 exhaust all the cases of NHS's with strictly 3-dimensional group of symmetries. 
All the other NHS's have either infinite dimensional symmetry groups or have symmetry groups of dimension no greater than 2.\\
\section{The classification of Section 4 and Appendix A}
Below we present a tree of the classification obtained in the preceeding sections.\\
\begin{picture}(500,278)
\thinlines
\drawframebox{104.0}{114.0}{60.0}{20.0}{}
\drawframebox{138.0}{214.0}{60.0}{20.0}{}
\drawframebox{268.0}{164.0}{60.0}{20.0}{}
\drawframebox{68.0}{164.0}{60.0}{20.0}{}
\drawframebox{202.0}{164.0}{60.0}{20.0}{}
\drawframebox{134.0}{164.0}{60.0}{20.0}{}
\drawframebox{208.0}{214.0}{60.0}{20.0}{}
\drawframebox{38.0}{114.0}{60.0}{20.0}{}
\drawframebox{172.0}{114.0}{60.0}{20.0}{}
\drawframebox{376.0}{114.0}{60.0}{20.0}{}
\drawframebox{308.0}{114.0}{60.0}{20.0}{}
\drawframebox{38.0}{54.0}{60.0}{40.0}{}
\drawframebox{204.0}{14.0}{60.0}{20.0}{}
\drawframebox{308.0}{64.0}{60.0}{20.0}{}
\drawframebox{376.0}{54.0}{60.0}{40.0}{}
\drawframebox{136.0}{14.0}{60.0}{20.0}{}
\drawframebox{172.0}{264.0}{60.0}{20.0}{}
\drawframebox{240.0}{114.0}{60.0}{20.0}{}
\drawframebox{240.0}{64.0}{60.0}{20.0}{}
\drawframebox{172.0}{64.0}{60.0}{20.0}{}
\drawframebox{104.0}{64.0}{60.0}{20.0}{}
\drawpath{172.0}{254.0}{138.0}{224.0}
\drawpath{172.0}{254.0}{208.0}{224.0}
\drawpath{138.0}{204.0}{68.0}{174.0}
\drawpath{240.0}{104.0}{38.0}{74.0}
\drawpath{240.0}{104.0}{104.0}{74.0}
\drawpath{138.0}{204.0}{134.0}{174.0}
\drawpath{208.0}{204.0}{202.0}{174.0}
\drawpath{208.0}{204.0}{268.0}{174.0}
\drawpath{68.0}{154.0}{38.0}{124.0}
\drawpath{68.0}{154.0}{104.0}{124.0}
\drawpath{202.0}{154.0}{172.0}{124.0}
\drawpath{202.0}{154.0}{240.0}{124.0}
\drawpath{268.0}{154.0}{308.0}{124.0}
\drawpath{268.0}{154.0}{376.0}{124.0}
\drawpath{308.0}{104.0}{240.0}{74.0}
\drawpath{308.0}{104.0}{172.0}{74.0}
\drawpath{376.0}{104.0}{308.0}{74.0}
\drawpath{376.0}{104.0}{376.0}{74.0}
\drawpath{172.0}{54.0}{136.0}{24.0}
\drawpath{172.0}{54.0}{204.0}{24.0}
\drawcenteredtext{172.0}{264.0}{{\tiny $\sigma=0$?}}
\drawcenteredtext{138.0}{214.0}{{\tiny $\rho=0$?}}
\drawcenteredtext{208.0}{214.0}{{\tiny $\rho=0$?}}
\drawcenteredtext{70.0}{164.0}{{\tiny $I=2c=$const?}}
\drawcenteredtext{134.0}{164.0}{{\tiny Proposition 3}}
\drawcenteredtext{202.0}{164.0}{{\tiny $J^0_1=\pm 1$? }}
\drawcenteredtext{268.0}{164.0}{{\tiny $I^0_1=$e$^{it}$?}}
\drawstackedtext{36.0}{114.0}{{\tiny $P=1+c\xi\bar{\xi}$ }\\{\tiny Proposition 2}}
\drawstackedtext{106.0}{114.0}{{\tiny $P=P(\xi,\bar{\xi})$ }\\{\tiny Proposition 2}}
\drawcenteredtext{172.0}{114.0}{{\tiny Proposition 10}}
\drawstackedtext{242.0}{114.0}{{\tiny are all $J^0_1$, $J^0_2$ $J^0_3$ }
\\{\tiny constants?}}
\drawcenteredtext{310.0}{114.0}{{\tiny $\der t\dz M\dz\bar{M}=0$? }}
\drawstackedtext{376.0}{114.0}{{\tiny are all $I^0_1$, $I^0_2$ $I^0_3$ }
\\{\tiny constants?}}
\drawstackedtext{38.0}{54.0}{{\tiny Bianchi types}\\{\tiny $VI_0,VII_0,$}
\\{\tiny $VIII,IX$}\\{\tiny Appendix A}}
\drawcenteredtext{106.0}{64.0}{{\tiny Proposition 9}}
\drawcenteredtext{172.0}{64.0}{{\tiny $t=\pi$?}}
\drawcenteredtext{242.0}{64.0}{{\tiny Proposition 6}}
\drawcenteredtext{310.0}{64.0}{{\tiny Propositon 4}}
\drawstackedtext{376.0}{54.0}{{\tiny Bianchi types}\\{\tiny $IV$, $VI_{-h}$, $VII_h$ }\\{\tiny Proposition 5}}
\drawcenteredtext{136.0}{14.0}{{\tiny Proposition 8}}
\drawcenteredtext{206.0}{14.0}{{\tiny Proposition 7}}
\drawcenteredtext{148.0}{240.0}{{\tiny Y}}
\drawcenteredtext{92.0}{190.0}{{\tiny Y}}
\drawcenteredtext{38.0}{140.0}{{\tiny Y}}
\drawcenteredtext{196.0}{190.0}{{\tiny Y}}
\drawcenteredtext{176.0}{140.0}{{\tiny Y}}
\drawcenteredtext{272.0}{140.0}{{\tiny Y}}
\drawcenteredtext{102.0}{90.0}{{\tiny Y}}
\drawcenteredtext{216.0}{90.0}{{\tiny Y}}
\drawcenteredtext{386.0}{90.0}{{\tiny Y}}
\drawcenteredtext{162.0}{90.0}{{\tiny N}}
\drawcenteredtext{136.0}{40.0}{{\tiny Y}}
\drawcenteredtext{200.0}{240.0}{{\tiny N}}
\drawcenteredtext{144.0}{190.0}{{\tiny N}}
\drawcenteredtext{248.0}{190.0}{{\tiny N}}
\drawcenteredtext{94.0}{140.0}{{\tiny N}}
\drawcenteredtext{228.0}{140.0}{{\tiny N}}
\drawcenteredtext{342.0}{140.0}{{\tiny N}}
\drawcenteredtext{264.0}{90.0}{{\tiny N}}
\drawcenteredtext{330.0}{90.0}{{\tiny N}}
\drawcenteredtext{200.0}{40.0}{{\tiny N}}
\end{picture}

\end{document}